\documentclass[12pt]{article}
\usepackage{graphics}
\newcommand{\la}{\label}

\newcommand{\be}{\begin{equation}}
\newcommand{\ee}{\end{equation}}
\newcommand{\ba}{\begin{eqnarray}}
\newcommand{\ea}{\end{eqnarray}}
\newcommand{\bastar}{\begin{eqnarray*}}
\newcommand{\eastar}{\end{eqnarray*}}

\title{Asymptotically Free Yang-Mills \\ \vskip 0.4cm
   Classical Mechanics with Self-Linked Orbits }
\author{ M.\ L\"ubcke\thanks{Martin.Lubcke@teorfys.uu.se },\hspace{1cm}
         A.J.\ Niemi\thanks{Antti.Niemi@teorfys.uu.se}\hspace{1cm}
         and\hspace{1cm}
         K.\ Torokoff\thanks{Kristel.Torokoff@teorfys.uu.se}\\ \\
         {\it Department of Theoretical Physics}\\
         {\it Uppsala University}\\
         {\it Box 803, SE-751 08 Uppsala, Sweden}}
\begin{document}

\maketitle

\begin{abstract}

\noindent
We construct a classical mechanics Hamiltonian
which exhibits spontaneous symmetry breaking akin
the Coleman-Weinberg mechanism, dimensional transmutation,
and asymptotically free self-similarity congruent
with the beta-function of four dimensional Yang-Mills theory.
Its classical equations of motion support
stable periodic orbits and in a
three dimensional projection these orbits are self-linked
into topologically nontrivial, toroidal knots.
\end{abstract}

%\end{titlepage}

The non-perturbative structure of four dimensional
Yang-Mills theory continues to be the subject of extensive
investigations. A major goal is the understanding of
large distance properties such as
color confinement, mass gap and the glueball spectrum.
The Yang-Mills theory has also a number of well
established salient features like ultraviolet
asymptotic freedom and the presence of finite action instantons.
Here we shall introduce a {\it classical} mechanics
Hamiltonian which contains many incredients of the
four dimensional Yang-Mills field theory, even though
it is defined in a four dimensional {\it phase} space.
These include asymptotically free self-similarity
with a coupling constant that flows like the one loop
coupling constant of four dimensional Yang-Mills theory,
dimensional transmutation, and spontaneous symmetry breaking
akin the Coleman-Weinberg
mechanism. Furthermore, we find that its Hamilton's
equations support stable periodic orbits.
Remarkably, we find that in a three dimensional projection
these orbits turn out to be self-linked into toroidal knots.
As usual, their self-linking number can then be computed by a
three dimensional Chern-Simons functional \cite{bott}.
This can be viewed as another trait of four dimensional
Yang-Mills theory.

We motivate our classical mechanics model by considering the infrared
four dimensional SU(2) Yang-Mills theory, in
a limit where all spatial inhomogeneities can be
ignored. In the maximal abelian gauge the classical
Yang-Mills action is then approximated by a classical
mechanics action \cite{savvidy}
\be
S \ = \ \int\limits_0^T d\tau \{ p_i \partial_\tau q_i -
\frac{1}{2}(p_1^2 + p_2^2) - \frac{1}{2} (q_1^2 - q_2^2)^2 \}
\la{sav}
\ee
where $\tau = t/t_0$ is dimensionless, and we have also
scaled $p_i$ and $q_i$ into dimensionless
quantities. When we rotate the coordinates by $\pi/4$ this becomes
the standard $x^2 y^2$ action \cite{simon}. This is a quite
universal model and besides the Yang-Mills theory it
relates {\it e.g.} to the low energy limit of
(super)membranes \cite{nico}. Despite its apparent simplicity
the $x^2y^2$ model has a number of remarkable
properties \cite{simon}. Most notably, the
classical dynamics of (\ref{sav}) is chaotic which originally
led to a conjecture that the model is ergodic. This was proven to be
wrong by \cite{dahl}, who found that the phase space
of (\ref{sav}) admits a non-ergodic island consisting
of a periodic orbit surrounded by a stable, invariant torus
of classical solutions. Since
the quantum mechanical partition function is evaluated
by the Gutzwiller trace formula \cite{gutz} which sums
over all periodic solutions to Hamilton's equations,
these stable periodic orbits are clearly
pivotal to the quantization of (\ref{sav}). In
fact, one could argue that if
(\ref{sav}) indeed does approximate the infrared limit of the
SU(2) Yang-Mills theory, the energies of these stable
periodic orbits are avatars of stable
glueball states in the Yang-Mills quantum theory.

In the present Letter we are interested in improving the relations
between (\ref{sav}) and the pure four dimensional SU(2) Yang-Mills
theory, maybe even the (super)membrane theory. Our starting point
is the one loop Yang-Mills effective action, originally
computed in \cite{savact}. Recently this computation has
been revisited in \cite{lisa}, employing a decomposed
\cite{fadde} Yang-Mills field. A comparison of
(\ref{sav}) with the results in \cite{lisa} suggests that we
introduce the following improvement,
\be
S \ = \ \int\limits_0^T d\tau \{ p_i \partial_\tau q_i -
\frac12 (p_1^2+p_2^2) - \frac12(q_1^2
-q_2^2)^2[1+\frac{\lambda}{2}\ln(q_1^2-q_2^2)^2]\}
\la{imph}
\ee
When $\lambda = 0$ we return to (\ref{sav}) and the potential
has a degenerate minimum with a vanishing energy $E=0$ along the
lines $q_1 = \pm q_2$.  For $\lambda > 0$
this minimum becomes unstable, there is a symmetry breaking akin the
Coleman-Weinberg mechanism and the new energy minimum occurs
along the four branches of
the hyperbola $|q_1^2 - q_2^2| = \exp(-\frac{1}{\lambda}
- \frac{1}{2})$. Consequently the new minimum
is also degenerate but with an additional four-fold
discrete multiplicity, and the new minimum energy is
\be
E_{min} = - \frac{\lambda}{4}
\exp(-1-\frac{2}{\lambda})
\la{mine}
\ee
Notice the non-analytic, non-perturbative
dependence on the coupling
in the small-$\lambda$ limit. We also note that
the minima of the $\lambda = 0$
potential energy of
(\ref{sav}) along the lines $q_1 = \pm q_2$ become local
maxima of the $\lambda > 0$ potential energy of (\ref{imph}),
separating the four branches of the $\lambda > 0 $
minima from each other. See figure 1.

We are interested in the periodic classical solutions to the
Hamilton's equations of (\ref{imph}), since these are the basic
incredients in the Gutzwiller trace formula quantization. We start
by revealing their self-similar scaling properties, in a manner
reminiscent of conventional renormalization group transformations.
For this we consider a periodic solution $\Gamma (\lambda,T)$ to
the Hamilton's equations, with some definite values of the period
$T$ and coupling $\lambda$. We then inquire how to scale this
periodic solution to other periods $T\to \tilde T$, possibly with
a redefinition of the coupling $\lambda \to \tilde \lambda$. For
this we introduce a change of variables in (\ref{imph}) which
scales $T \to \tilde T$, redefines $\lambda \to \bar \lambda$ and
possibly multiplies the action by some overall constant $S \to
\kappa S$ but leaves the functional form of the integrand in
(\ref{imph}) otherwise intact. The form of Hamilton's equations
remains then untouched but a periodic orbit $\Gamma(\lambda,T)$
becomes mapped into a new periodic orbit $\Gamma(\bar\lambda ,
\bar T)$ solving the original equations with the new parameter
values $\bar \lambda, \ \bar T$.

To implement this scaling transformation explicitly, we first
consider the scaling of periodic trajectories in the $x^2y^2$
model (\ref{sav}),
\be
\begin{array}{ccc}
t & \longrightarrow & c^{-1} t \\
q & \longrightarrow & c q \\
p & \longrightarrow & c^{2} p
\end{array}
\label{scale1}
\ee
This transformation leaves the equations of
motion in (\ref{sav}) invariant, and trajectories with period $T$
are mapped into trajectories with period $c^{-1}T$.

In the case of (\ref{imph}), the scaling (\ref{scale1}) fails to
leave the functional form of the equations of motion intact. For
this, we need to modify the scaling of $t$ and $p$ as follows:
\be
\begin{array}{ccc}
t & \longrightarrow & \left( \frac{1 -
\lambda \ln c^2}{c^2}\right)^{\frac{1}{2}} t \\
p & \longrightarrow & \left( \frac{c^4}{1-\lambda \ln
c^2} \right)^{\frac{1}{2}} p
\end{array}
\ee
and in addition we must redefine
\be
\begin{array}{ccc}
\lambda & \longrightarrow & \frac{\lambda}{1-\lambda \ln c^2}
\la{beta}
\end{array}
\ee
 This is then a self-similarity transformation which leaves the overall 
functional
form of the Hamilton's equations of (\ref{imph}) intact;
The only
effect of this improved scaling transformation in (\ref{imph}) is
a renormalization of the coupling constant $\lambda$ according to
(\ref{beta}), a scaling of period $T$ and an overall
multiplicative redefinition of the action.

Note that the coupling constant flow (\ref{beta})
is like the flow of the one-loop coupling constant
in four dimensional Yang-Mills theory. In particular, in
the limit of small $c$ the periodic classical solutions of
(\ref{imph}) exhibit asymptotically free self-similarity,
in the sense that in this limit (\ref{imph}) approaches
(\ref{sav}).

When we increase $c$ we find an upper bound $c^2_{max} = \exp(1/\lambda)$,
where the renormalized coupling constant (\ref{beta}) diverges.
In the present case this Landau pole is physically relevant.
It determines a critical value of $c$ which sets a
lower bound for the period $T_{min}$, below which a periodic
orbit can not be extended by the self-similarity
transformation.

The Hamilton's equations of (\ref{imph}) can be readily solved
by numerical integrations. Previously it has been shown \cite{dahl}
that with $\lambda = 0$ there is a periodic solution which is
surrounded by an stable, invariant torus formed by classical
solutions. Subsequently the existence of additional
periodic solutions with accompanying invariant torii
has been reported {\it e.g.} in \cite{dahl2}.
We have investigated the effects of a nonvanishing $\lambda$ in
(\ref{imph}) to the stability of the solution presented in \cite{dahl}.
When we increase $\lambda$ from $\lambda = 0$
while keeping the total energy of the
solution intact, we find that the torus of classical solutions
which surrounds the periodic orbit of \cite{dahl} retains its
invariant character for small values of the coupling $\lambda$.
But when the coupling approaches a critical value $\lambda_c
\approx 0.6$ we find that the torus looses its invariance properties
essentially by period doubling. This indicates that at $\lambda_c$
the periodic orbit of \cite{dahl} looses its stability.

However, for non-vanishing $\lambda$ we also find {\it novel}
periodic solutions with the physically interesting property
that their energies are negative, $E < 0$. Consequently
these solutions have an energy which is {\it lower} than that
of the $\lambda = 0$ ground state. These solutions are also
surrounded by stability islands which are formed by invariant
torii of $E < 0$ (not necessarily periodic) classical solutions.
Since $E < 0 $ these stability islands consist of
trajectories which are entirely confined in one of the four
energy valleys which surround the minimum energy hyperbola
$|q_1^2 - q_2^2| = \exp(-\frac{1}{\lambda} - \frac{1}{2})$.
Consequently we can describe their properties by restricting
to one of the four valleys, and for definiteness we shall select the
valley where $q_1>0$. We also employ the self-similar
scaling property of the action (\ref{imph}) to set
$\lambda = 1$. This is akin a dimensional transmutation,
where instead of $\lambda$ the solutions are characterized by
some other distinguishing parameter. For this parameter
we choose the maximal value of $q_1$ which is attainable to the
trajectory, say at $q_2 = 0$. Notice that since the
energy is the sole conserved
quantity in (\ref{imph}), the actual motion described by a
classical solution occurs in a three dimensional space.
Consequently we can visualize the trajectories in terms of
a three dimensional projection, and for this we select the
subspace $(q_2,p_1,p_2)$. In particular, we can characterize
the trajectories using invariants of this three dimensional
space such as their self-linking number, provided the trajectories
are indeed linked.

By numerical integration we find that in the $q_1 > 0$
valley the Hamilton's equations of (\ref{imph}) describe a periodic
orbit which is stable (when $\lambda = 1$) in the interval $E_{min}
\approx -0.0045 \leq E \leq E_{max} \approx -0.0008$. In figure 2
we draw the solution for $E = -0.0010$ in the $(q_1 , q_2)$ plane.
In figure 3 we plot the Poincare map of the stable solution and
some trajectories in the surrounding invariant torus at the
$q_2 = 0$ surface of section; The stability of the solution and
its invariant torus is evident.
In figure 4 we describe how this solution scales, by plotting the
the energy of the trajectory as a function of the maximal value
of $q_1$; Notice that by dimensional transmutation
this corresponds to changing $\lambda$ in the
original model.

We find that the
periodic orbit still exists when the energy $E$
of the trajectory exceeds a critical value $E_{max}
\approx -0.0008$. But when $E$ exceeds $ E_{max}$
the solutions in the torus loose
their stability and the torus shrinks away. At that point
our periodic orbit looses its stability.

Finally, we have performed a detailed investigation of the solutions
that form the invariant torus. For rational windings around the
torus, these solutions are themselves closed orbits. Their
geometrical shape can be visualized in the three dimensional
$(q_2,p_1,p_2)$ subspace, where the orbits
become closed curves in $R^3$. Remarkably, we find that
these curves can be self-linked into topologically nontrivial
torus knots. As an example, in figure 5 we describe a
closed orbit which we identify as a figure-eight knot in
the three dimensional subspace. Indeed, we propose that
{\it all} torus knot can appear as stable closed orbit solutions
to the Hamilton's equations of (\ref{imph}). This is
reminiscent of the knots \cite{nature} in a model that
relates to the four dimensional Yang-Mills theory \cite{fadde}.

\vskip 0.5cm
In conclusion, we have studied a classical mechanics model which
relates both to the four dimensional Yang-Mills theory and the
(super)membrane. We have found that its periodic classical solutions
exhibit self-similarity with a coupling
constant flowing like the Yang-Mills coupling constant.
Surprisingly, we have found that such periodic
solutions exist even for energies which are lower than the energy
of the ground state at vanishing coupling. Furthermore, these
solutions are stable in the sense that they are surrounded by
an invariant torus which is
formed by classical solutions. In this torus we then
identify several additional periodic solutions, some of which form
nontrivial torus knots. If our model indeed relates to the low energy
limit of four dimensional Yang-Mills theory these solutions
could be the avatars of its glueball states.

\vfill\eject

\vfill\eject

\begin{figure}
{\centering \resizebox*{6cm}{4cm}
{\rotatebox{-90}{\includegraphics{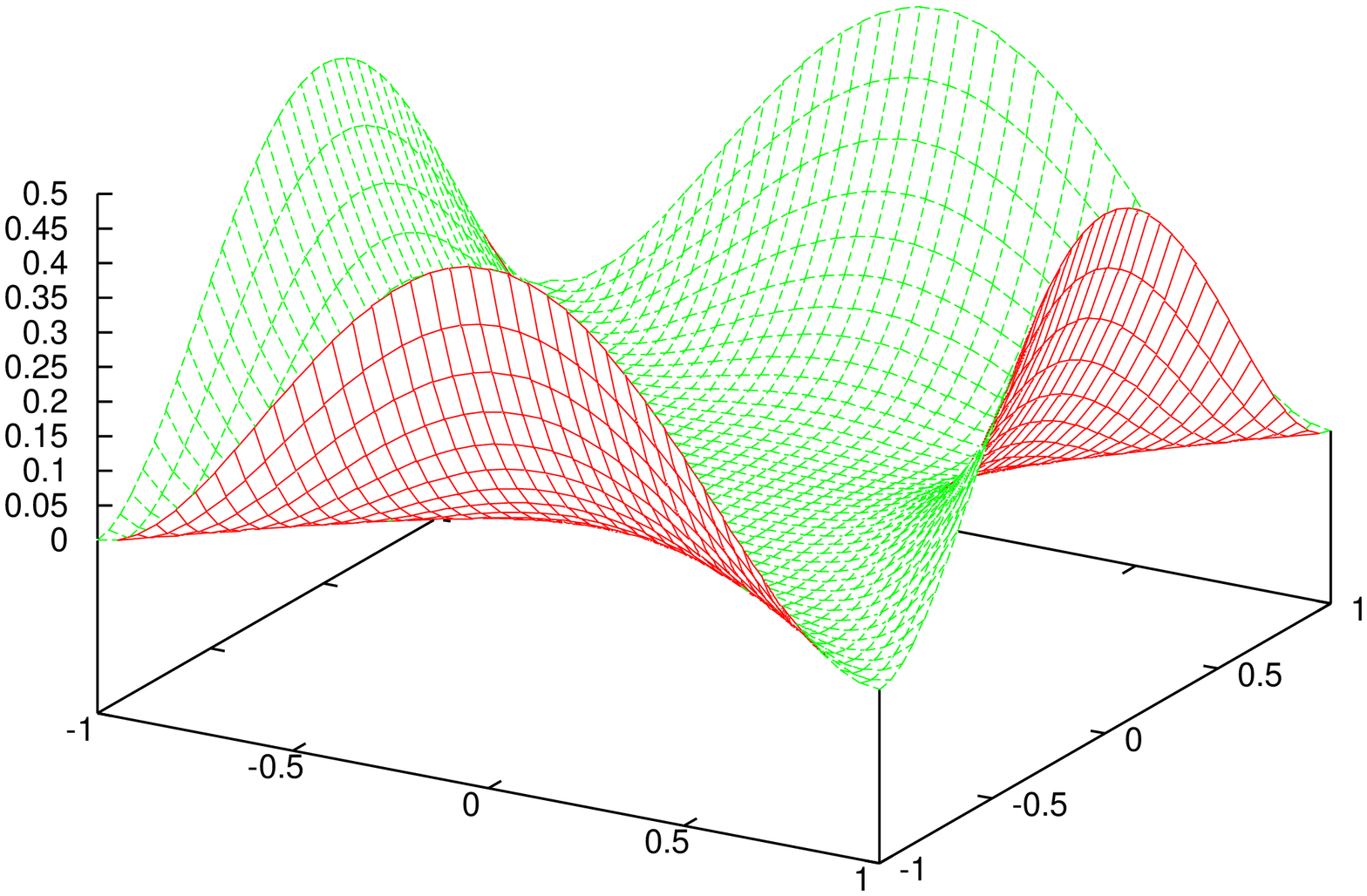}}} \resizebox*{6cm}{4cm}
{\rotatebox{-90}{\includegraphics{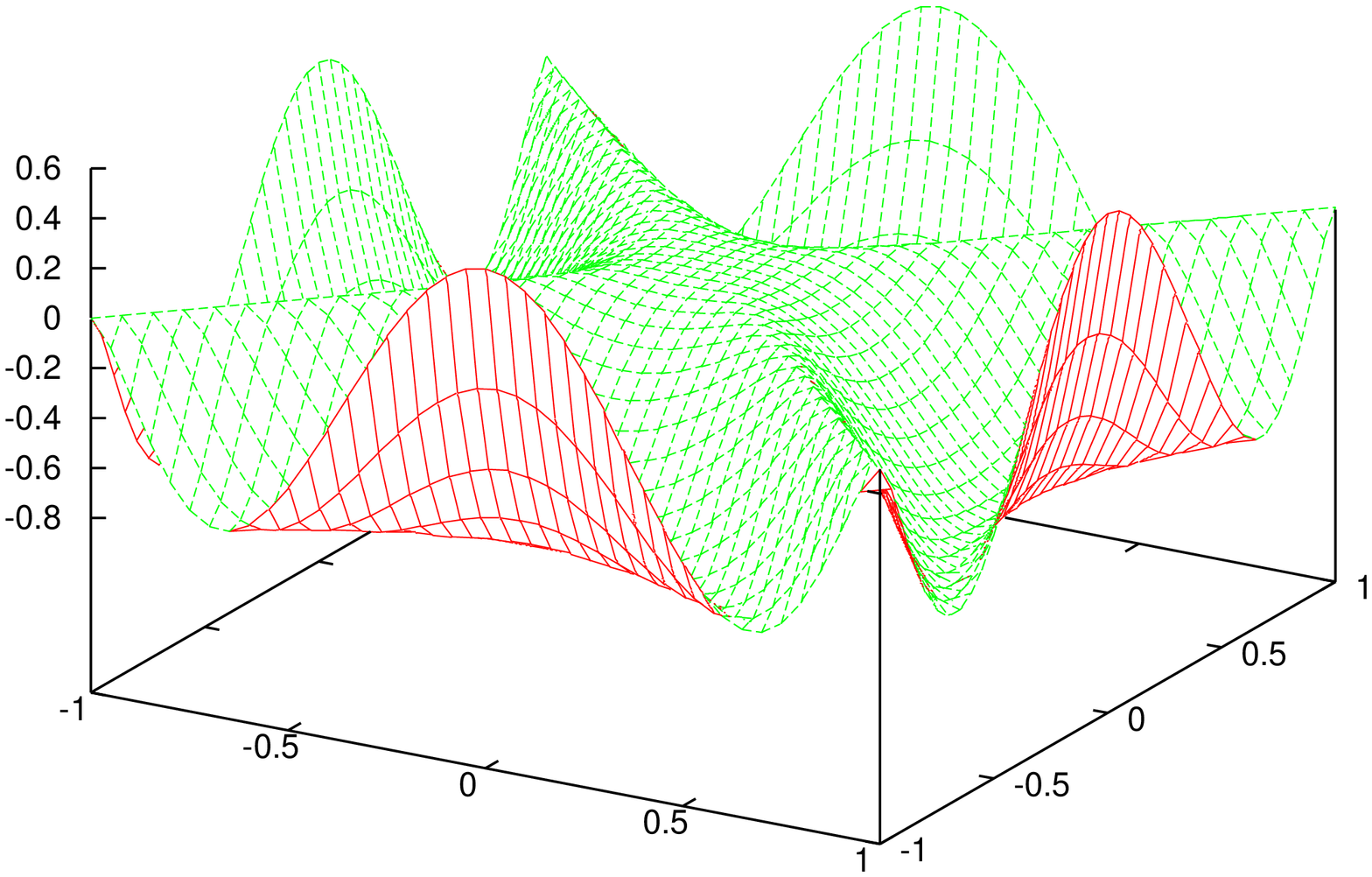}}} \par} \caption{The
potential in (\ref{imph}) for $\lambda=0$ and $\lambda=10$
respectively.}
\end{figure}

\begin{figure}
{\centering \resizebox*{8cm}{8cm}
{{\includegraphics{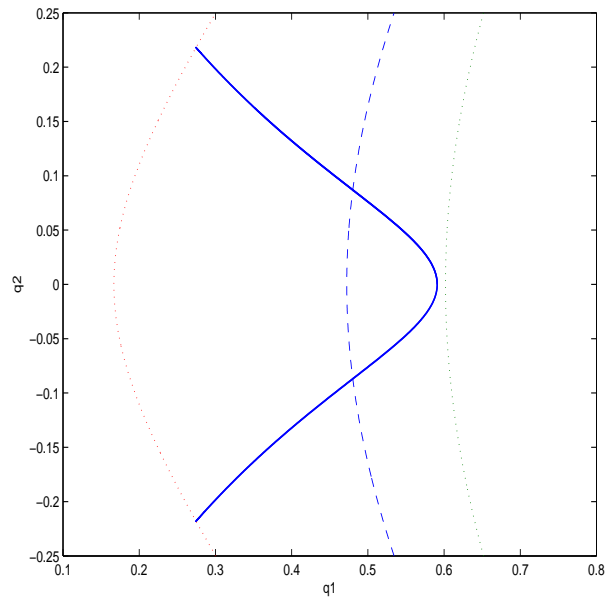}}} \par} \caption{The stable
trajectory at $\lambda_1$ and $E=-0.001$, drawn on the $(q_1,
q_2)$ plane. The dotted lines denote where the potential is
$V=-0.001$ and the dashed line is the minimum of the potential.}
\end{figure}

\begin{figure}
{\centering \resizebox*{8cm}{8cm}
{\rotatebox{-90}{\includegraphics{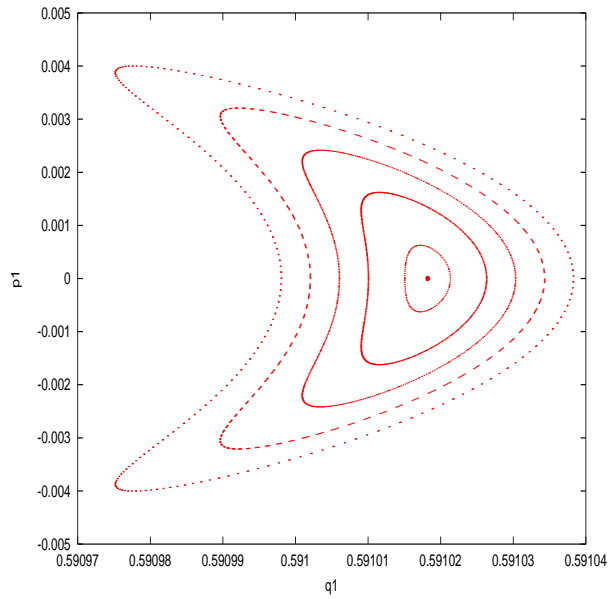}}} \par}
\caption{Poincare section at $q_2=0$ which describes the stable
solution at the center and selected trajectories of the
surrounding invariant torus.}
\end{figure}

\begin{figure}
{\centering \resizebox*{8cm}{8cm}
{\rotatebox{-90}{\includegraphics{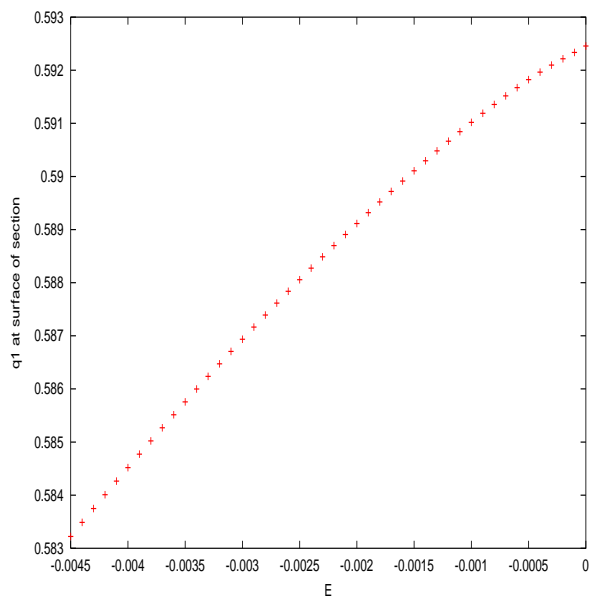}}} \par} \caption{The
relation between energy and maximal value of $q_1$, at $q_2=0$.}
\end{figure}

\begin{figure}
{\centering \resizebox*{8cm}{8cm}
{{\includegraphics{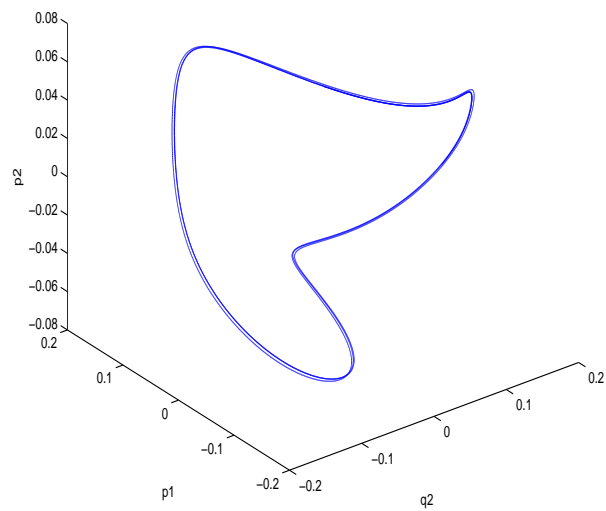}}}
\par} \caption{The figure-8 solution in the $(q_2,p_1,p_2)$
subspace. Notice that the torus around which the
solution wraps has a very small radius in comparison to its
length. }
\end{figure}

\end{document}